\documentclass{F99UXus}
\usepackage{epsfig}

\newcommand{\bit}{\begin{Itemize}}
\newcommand{\eit}{\end{Itemize}}

\begin{document}
\title{
\small{{\em Presented at $e^+e^-$ Factories 1999, Sept. 21-24, 1999,
KEK Laboratory} \hfill CBN 99-30} \\
\vspace*{0.2in}
{\Large\bf Beam-Generated Detector Backgrounds at CESR}
\thanks{Work supported by the National Science Foundation.}
}
\author{Stuart Henderson
\thanks{Email: sdh9@cornell.edu},
Laboratory of Nuclear Studies, Cornell University, Ithaca NY, 14853, USA \\
David Cinabro, Wayne State University, Detroit MI, 48202, USA
}

\maketitle

\begin{abstract}
The CESR/CLEO Phase II interaction region is described.  The operational
experience with beam-generated detector backgrounds is reviewed.  The
status of our understanding of beam-generated detector backgrounds at CESR
is described and comparisons of background measurements with simulation
predictions are presented.
\end{abstract}

\section{INTRODUCTION}
A new CESR/CLEO interaction region 
\cite{ref:henderson_pac97,ref:henderson_hawaii}
was installed in 1995 as part of the
CESR Phase II upgrade.  At the same time the CLEO experiment
\cite{ref:cleo} installed a 
silicon vertex detector (SVX) around a 2 cm radius beryllium beampipe,
and new detector shielding.  (The new configuration of the CLEO detector
is referred to as CLEO II.V\cite{ref:cleosvx})
The use of
radiation sensitive SVX readout electronics located just outside the
central beampipe placed a premium on
accurate prediction and understanding of beam-related detector backgrounds
at CESR.  This article describes the detector shielding system,
commissioning and operational experience
and presents a status report on our present understanding of
beam-generated
detector backgrounds at CESR.

The CESR Phase II interation region (IR) was optimized for bunch train 
collisions at a small horizontal crossing angle (see \cite{ref:henderson_kek99}
in these proceedings).  CESR presently operates with 9 trains of bunches, each
of which consists of 4 bunches spaced by 14 ns.  The peak stored beam
current to date is 550 mA with nearly equal electron and positron beam
currents.

The dominant CLEO detector background arises from lost particles
generated by bremsstrahlung interactions with the residual gas within
about 30 m of the IP.
Coulomb scattering is a small contribution as is the dose from 
scattered synchrotron radiation.  The CLEO experiment is
sensitive to beam-generated backgrounds through the following mechanisms:
i) background hits (detector occupancy) in the drift chambers,
ii) background hits (occupancy) in the SVX,
iii) radiation dose to the SVX readout electronics, iv) radiation
dose to the CsI crystal calorimeter and v) increased data-acquisition
trigger rates.
It turns out that the 
radiation sensitivity of the silicon readout electronics 
(CAMEX amplifiers) placed
the most stringent requirements on the maximum tolerable
detector background rates.  Radiation damage studies \cite{ref:camex}
found that the failure dose for the powered CAMEX amplifiers is
$\sim$25 krad ($\sim$100 krad unpowered).
For detector shielding design purposes we chose the following design criteria:
i) SVX layer 1 occupancy $< 1$\% hits/strip/$\mu$s, 
and ii) SVX readout electronics dose $< 20$ krad in 3 years.

\section{CESR Phase II Interaction Region}

\subsection{Design Considerations}

The crossing angle orbit has important consequences 
for synchrotron radiation backgrounds.  The incoming beam
is off-axis in the interaction region quadrupoles, perhaps by as much as 
1 cm, generating substantial SR power
only a few meters from the IP.  This source alone would lead to 
unacceptably large detector backgrounds.
In order to bring the incoming beam nearer the 
axis of the nearest horizontally 
focusing quadrupole, the beam is displaced horizontally with a magnetic
IP displacement bump.  This bump was included in the lattice 
design phase to ensure
adequate aperture.  With the use of this bump, SR powers are reduced and 
with proper shielding the background rates are acceptable.  In addition, for
injection the beams are separated horizontally at the IP,
again placing one beam off-axis in the IR quads and producing higher SR
fluxes during injection than HEP.

The detector background shielding system was designed based on the
results of detector background simulations which were developed for
lost particles and synchrotron radiation.  These simulations are
described in detail elsewhere\cite{ref:henderson_hawaii, ref:cesrb}.
The results of these simulations were compared to detector background
measurements performed with CLEO-II and were found to be in very good
agreement providing confidence in their predictive power
\cite{ref:cinabro_hawaii, ref:cinabro_glasgow, ref:cinabro_dpf,
ref:cesrb}.

\subsection{Detector Shielding}

The detector shielding consists of two parts.  The detector is shielded from
lost beam particles by a large tungsten mask approximately 2.5 cm thick
extending from $\sim$24 cm to $\sim$50 cm from the IP.  In addition,
the rare-earth permanent magnet quadrupole material provides 
additional shielding outboard of the tungsten mask.  Tungsten was selected
as the mask material because of its short radiation length.  The detailed shape
was designed based on the results of Monte-Carlo simulations of beam-gas 
scattering.  The detector is shielded from synchrotron radiation by
a stepped copper mask.  The innermost mask tip is located at
$r=1.2$ cm and $z=26$ cm.  The mask is profiled in such a way that SR photons
strike only the tips of the mask structure;  the surfaces parallel to the
beam direction are angled slightly so that these surfaces intercept no
SR flux that would otherwise forward scatter into the detector beampipe.
As a result, all incoming SR power is absorbed on the tips.
Rather than make these tip surfaces at right angles to the beam direction,
the surfaces are sloped to reduce the HOM power generated in the structure.
The slope is small enough that the tip surfaces remain hidden from the
detector beampipe.
Thus, the background from SR arises from photons which scatter through
a mask tip (``tipscattering'') and photons which pass over the innermost
mask tip and strike the downstream surface of the mask and subsequently
scatter into the beampipe (``backscattering''). 

The central detector beampipe \cite{ref:henderson_beampipe} 
is a double-walled water cooled beryllium
beampipe with inner radius $r=1.9$ cm.
To further reduce the SR backgrounds a $10$ $\mu$m gold coating was 
deposited \cite{ref:henderson_gold} on the
inner surface of the beampipe.  
The coating reduces the radiation dose to the SVX detector by 3-4 
orders of magnitude.
To further reduce the lost 
particle background the innermost mask tip is made of
tantalum rather than copper.
Since during injection one of the beams is off-axis in the IR quads, the 
possibility for large SR fluxes exists.  A set of retractable 
tungsten foil radiation shields
are mounted to the beampipe.  These tungsten shields were 
closed during CESR injection for the first year of operation, and
then were left open during HEP operations and closed during
machine studies.

In order to reduce beam-gas scattering near the IR new vacuum
chambers and pumping was installed within $\pm$12 m of the IP.
IR Pumping is accomplished in large plenums which incorporate
massive titanium sublimation pumping
\cite{ref:nbm_pumping}.  Pumping plenums are located at
$\pm$ 2m, $\pm$ 6 m, and $\pm$ 10 m from the IP.

Detector backgrounds are monitored with two systems.  The first is a
PIN diode radiation monitoring system mounted 
directly on the ends of the beryllium beampipe.  Twelve diodes are
mounted on each end and are uniformly distributed in azimuth.  This
system is sensitive to both charged particles and SR photons.  Another
system, consisting of a set of CsI 
crystals placed in the CLEO endcaps, is useful
for monitoring charged particle backgrounds.

The background simulation predictions for
design currents (600 mA total) and a 40\% duty factor 
are given in Table \ref{table:simulation}.  
Both the expected dose and occupancy are below the
design goals mentioned above.  As can be seen the beam-gas background
dominates.
\begin{table}[htb]
\begin{center}
\caption{Detector background simulation results for 600 mA total 
current and a 40\% duty factor.}
\medskip
\begin{tabular}{|c|c|c|}
\hline
   &  SVX layer 1 Occ. & Dose \\
Source  & \% hits/layer/$\mu$s & krad/year \\ \hline
SR & 0.002 & 0.64 \\
Beam-Gas & 0.18 & 8 \\ \hline
Total & 0.18 & 9 \\ \hline
\end{tabular}
\label{table:simulation}
\end{center}
\end{table}

\section{Operational Experience}

\subsection{Total Dose and Average Rates}

\begin{figure}[tbh]
\epsfig{bbllx=40,bblly=150,bburx=530,bbury=620,file=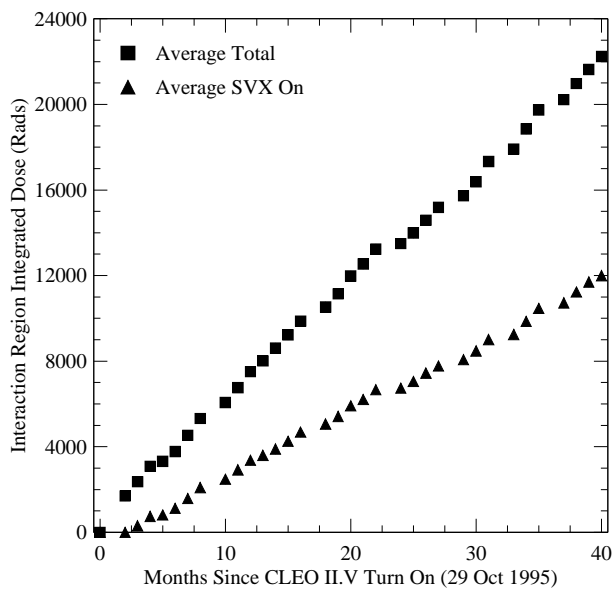, width=82.5mm}
\caption{The integrated interaction point dose versus month
of CLEO II.V running.}\label{fig:dose}
\end{figure}

	CLEO~II.V ran for a total of 172 weeks from 29 October 1995
until 15 February 1999.  The first month was completely devoted to machine
studies, the next two months to a detector engineering run,
but after that the detector ran continously for three years.
There were eight scheduled maintenance periods lasting, on average, three
and a half weeks and three unscheduled periods each of less than one week.
The integrated running
time of the detector was 142 weeks.  During that time the total 
accumulated radiation dose measured by the beampipe PIN diode monitors
was 22.2 krads.
Of that, 14.8 krads were logged when  
CESR was delivering luminosity to CLEO and 12.0 krads when
CLEO was actually taking data.  The dose when CESR was not
delivering luminosity was dominated by injection, and amounts to
3.6 krads.  This implies that injection for luminosity represents
30\% of the dose logged in data taking.  
The balance of the dose was logged in machine
studies (3.5 krads) and during the short time
spent between injection and HEP data-taking (0.3 krads).
Most of the dose during machine studies ($\sim$75\%)
was logged during injection.
The integrated dose versus month of running is shown in
Figure~\ref{fig:dose}.

	During the course of the CLEO II.V run CESR delivered 11.6 fb$^{-1}$
of luminosity and CLEO logged 9.0 fb$^{-1}$.  
%The integrated luminosity versus
%month of CLEO II.V running is shown in Figure~\ref{fig:lum}.
%\begin{figure}
%\epsfig{bbllx=25,bblly=120,bburx=530,bbury=700,file=cleo25_int.ps, width=82.5mm}
%\caption{The integrated luminosity logged versus month
%of CLEO II.V running.}\label{fig:lum}
%\end{figure}
The total Amp-Hours accumulated during the CLEO II.V run
was just over 4000.  
%Figure~\ref{fig:amphrs} shows the accumulated
%\begin{figure}
%\epsfig{bbllx=40,bblly=150,bburx=530,bbury=620,file=amphrs.ps, width=82.5mm}
%\caption{The accumulated Amp-Hours in CESR versus month
%of CLEO II.V running.}\label{fig:amphrs}
%\end{figure}
%Amp-Hours versus month.
These imply that the total dose rates were
0.16 krads/running week, 1.9 krads/fb$^{-1}$ delivered,
2.5 krads/fb$^{-1}$ logged,
and 5.6 rads/Amp-Hour. Based on these
numbers and the expected radiation hardness of the SVX
readout electronics
the detector could have survived for 30\% longer.
We did have one readout chip fail in the way expected from radiation
damage; a sudden increase in noise in all the channels in the chip 
was observed
in the last six months of running.  This chip was in layer 3 where
lower quality chips were used, but the dose rate was expected to
be more than a factor of two smaller than that observed by
the PIN diode system which is located near layer 1.

During the Phase II IR commissioning period, 2.4 krads were accumulated in
the first three months with only 155 A-hr.  There are several reasons for 
this high specific dose rate.  The dynamic pressure rise ($dP/dI$)
was naturally large during this period since new vacuum chambers
were installed in the IR.  In addition, much of the time was spent
tuning high current injection which has a higher dose rate than
colliding beam running.  This startup period represented a dose/A-hr
about three times what was achieved in normal running.

%	The start-up period integrated a total dose of 2.4 krads
%in three months and only 155 Amp-Hours and no useful
%data logged.  There are many reasons for this large initial dose rate
%per Amp-Hour;
%about three times the average.  The initial month of machine studies
%was dominated by injection studies which
%are a higher rate of dose than luminosity running.
%Also the pressure per unit beam current was on average twice that of the
%value it eventually stabilized.  Our start-up period represented
%a dose about three times what we achieved in normal running in terms
%of Amp-Hours.  In terms of time is was nearly normal as during
%the start-up period beam currents were limited as a learning
%curve on high beam current injection was climbed
%and by a hard limit on the dose rate observed on the PIN diode system
%which would dump the beams by turning off the power to CESR's RF system.

\section{Comparison of Measurements with Simulation}

\subsection{Pressure Bump Studies}
\begin{figure}[tbh]
\epsfig{bbllx=40,bblly=150,bburx=530,bbury=635,file=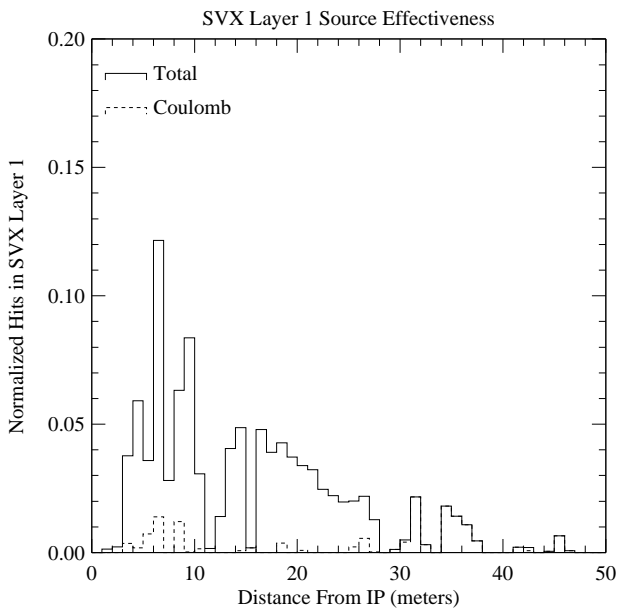,width=82.5mm}
\caption{The source effectiveness distribution.
This plot shows the relative contribution to the SVX layer 1 occupancy from
beam-gas scattering as a function of scatter (source) position.
}\label{fig:source}
\end{figure}

Controlled pressure bump studies are among the most important tests of
the beam-gas background simulation since such tests check directly the
model of the source-effectiveness for scattered beam particles.
By source-effectiveness we mean the relative contibution to detector
background due to beam-gas scattering at a particular location in
the ring.  The source-effectiveness for SVX layer 1 hits predicted
by the beam-gas simulation is shown in Figure~\ref{fig:source}.
(Reference \cite{ref:yulin_kek99} in these proceedings 
shows another source effectiveness
calculation, that for IR mask strikes.)  The actual background
is therefore the source-effectiveness weighted by the pressure
profile scaled appropriately by the beam current.

\begin{figure}[tbh]
\epsfig{bbllx=40,bblly=150,bburx=530,bbury=635,file=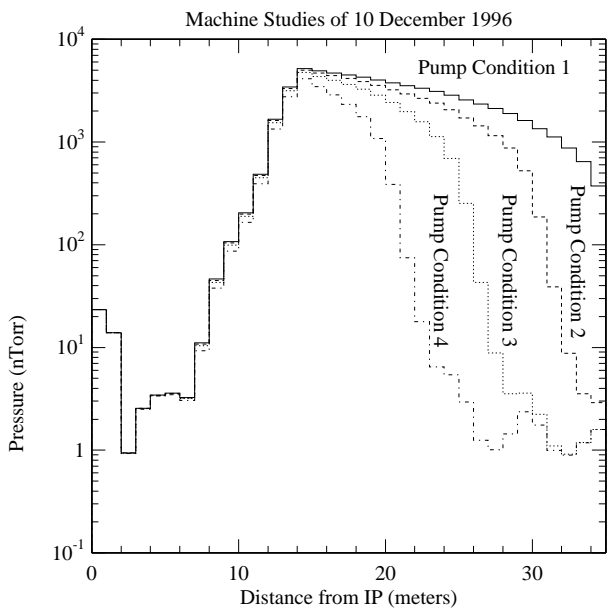,width=82.5mm}
\caption{Pressure distribution near the CESR/CLEO II.V interaction region
during a controlled leak pressure bump study for various
vacuum pump conditions.}\label{fig:press1}
\end{figure}

\begin{figure}[tbh]
\epsfig{bbllx=40,bblly=150,bburx=530,bbury=635,file=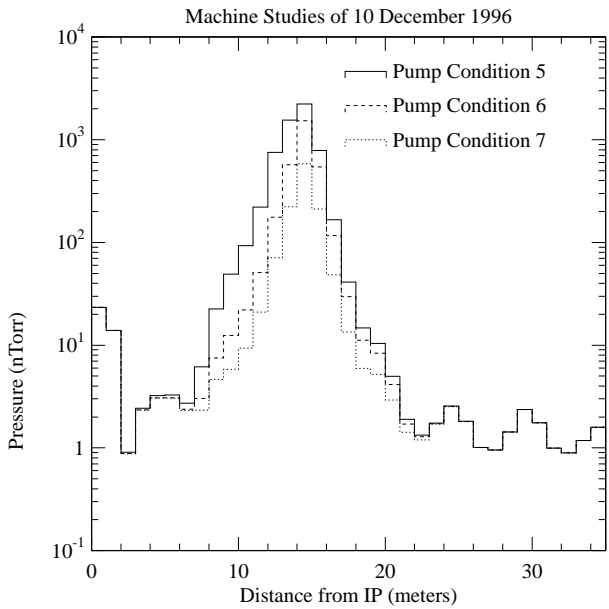,width=82.5mm}
\caption{Pressure distribution near the CESR/CLEO II.V interaction region
during a controlled leak pressure bump study for various 
vacuum pump conditions.}\label{fig:press2}
\end{figure}

\begin{figure}[tbh]
\epsfig{bbllx=40,bblly=150,bburx=530,bbury=635,file=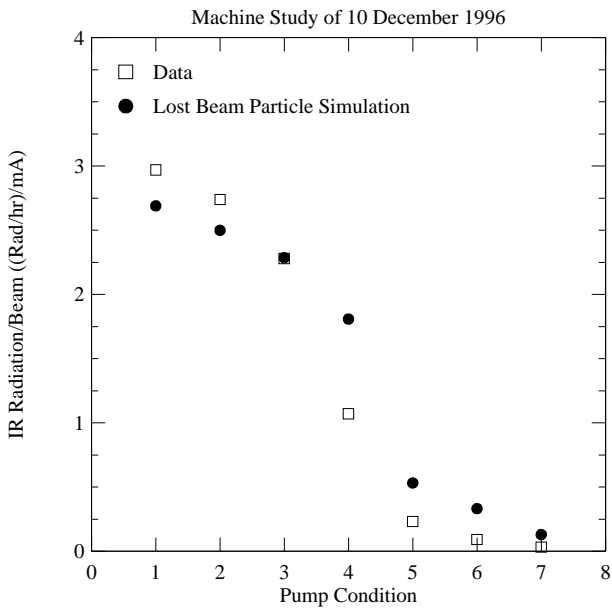,width=82.5mm}
\caption{Radiation dose rate at the CESR/CLEO II.V IP
for the seven pump conditions shown in Figures~\ref{fig:press1}
and~\ref{fig:press2} compared with the 
prediction of the beam-gas simulation.}\label{fig:radleak}
\end{figure}

We have performed such studies by turning off pumps near the IR to
generate pressure bumps and by introducing calibrated leaks.
We report on one such study in which a calibrated N$_2$ leak was
introduced 12 m from the IP.  A series of ``pump conditions'' was
generated by turning off successive pumps near the IR
\cite{ref:nbm_con}.  Pressures
were recorded for each pump condition.  These data were used together
with a model of pumping in the IR (described in \cite{ref:yulin_kek99}
in these proceedings) to generate a predicted pressure profile.
Pressure profiles for seven pump conditions predicted by the model are
shown in Figures \ref{fig:press1} and \ref{fig:press2}.
The detector radiation dose was measured for each of these seven conditions.
The pressure profiles together with the source-effectiveness and
known beam current provide a simulation prediction of the radition dose
for each of the pump conditions.  The comparison is shown in Figure
\ref{fig:radleak}.  We take the good agreement as confirmation that the
beam-gas source-effectiveness predicted by the simulation is correct.

%
%	Our simulations of beam generated backgrounds in CLEO\cite{simu}
%have been checked with numerous machine studies of the course of the
%CLEO~II.V experiment.  The most important check of the background
%generated by beam particles lost as they scatter off the residual
%gas in the rings  are controlled pressure
%bump studies.  These check if our model for the source effectiveness for
%scattered beam particles to create backgrounds in CLEO is correct.  These
%studies are done by opening calibrated leaks of nitrogen gas into the ring
%at various points near the interaction point (IP).  During the course of the
%controlled leak the pressure is carefully
%monitored in the region near the leak.  These studies are also extremely
%valuable in checking our models of the pressure distribution around
%the ring as they give data on pumping speeds and gas conductances for 
%the CESR vacuum system.\cite{nari+yulin}  
%When the pressure has stabilized with the controlled
%leak beam is placed in the machine and backgrounds are measured.  The pressure
%distribution can be altered by turning on and off vacuum pumps and many
%experiments can be carried out.  Figures~\ref{fig:press1} and~\ref{fig:press2}
%show the calculated pressure distribution for seven different pumping
%conditions for a calibrated leak opened twelve meters from the CLEO IP.  
%Figure~\ref{fig:radleak} shows a comparison between the measured radiation
%rate at the CLEO IP and the simulation's predicted rate.  The agreement
%is good and we take this as a confirmation of the simulations prediction
%of the source effectiveness which is shows in Figure~\ref{fig:source}.
%Note that Figure~\ref{fig:source} is for occupancy in the first layer
%of the CLEO~II.V Silicon Vertex Detector (SVX).  The source effectiveness
%for radiation on the IP and outer layers of CLEO~II.V tracking system
%are similar.

\subsection{High Energy Physics Running}

	We can also check the prediction of our simulation in standard high
energy physics (HEP) running conditions.  This prediction depends on the
pressure
distribution which is taken from a calculation\cite{ref:yulin_private} 
done at
\begin{figure}[tbh]
\epsfig{bbllx=40,bblly=150,bburx=530,bbury=635,file=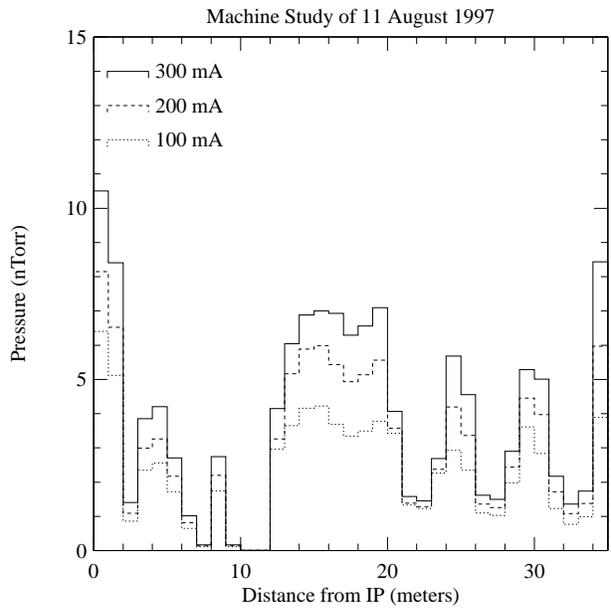,width=82.5mm}
\caption{The calculated pressure profile near the CESR/CLEO~II.V IR
during HEP running.}\label{fig:HEPpress}
\end{figure}
\begin{figure}[tbh]
\epsfig{bbllx=40,bblly=150,bburx=530,bbury=635,file=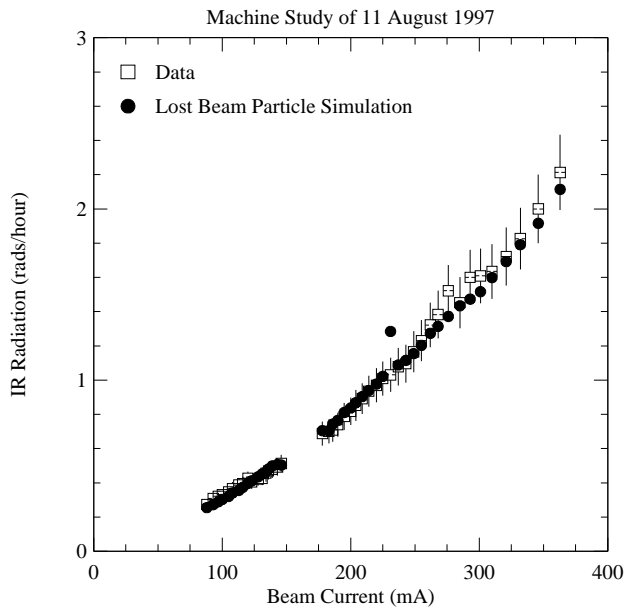,width=82.5mm}
\caption{The radiation level at the CESR/CLEO~II.V IP
during HEP running.  The prediction of the lost beam particle simulation
and the measured data are shown.}\label{fig:HEPrad}
\end{figure}
a beam current of 300 mA.  The extrapolation to different beam currents
is done by assigning 
the pressure measurements (cold-cathode gauges or ion pump currents)
to regions of the beamline and appropriately scaling with beam current.
The pressure distributions used by the simulation for HEP running are shown in
Figure~\ref{fig:HEPpress} for various beam currents.  The pressure distribution
is combined with the source effectiveness and beam current to give the
background levels as a function of beam current.  Figure~\ref{fig:HEPrad}
shows the radiation level at the IP.  The agreement is striking.

\begin{figure}[tbh]
\epsfig{bbllx=40,bblly=150,bburx=530,bbury=635,file=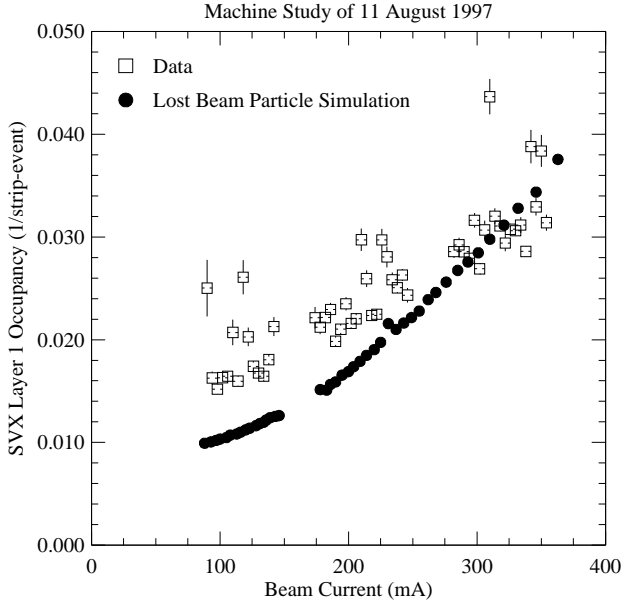,width=82.5mm}
\caption{The occupancy in the first layer of the CLEO~II.V SVX
during HEP running.  The prediction of the lost beam particle simulation
and the measured data are shown.}\label{fig:HEPs1}
\end{figure}

\begin{figure}[tbh]
\epsfig{bbllx=40,bblly=150,bburx=530,bbury=635,file=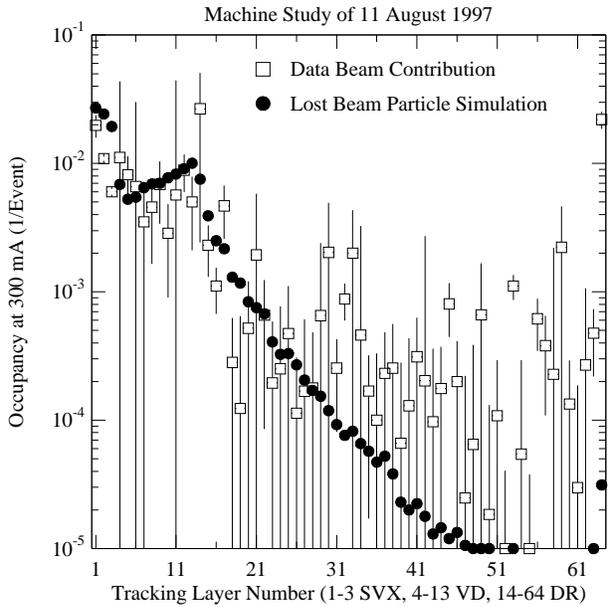,width=82.5mm}
\caption{The occupancy as a function of tracking layer number in CLEO~II.V
during HEP running.  The prediction of the lost beam particle simulation
and the measured data are shown.}\label{fig:HEPovsl}
\end{figure}

	Since HEP data was also taken during this test we can compare the 
measured and predicted
occupancy levels in the CLEO~II.V detector.
The measured and predicted SVX layer 1 occupancy are shown in 
Figure~\ref{fig:HEPs1}.  The occupancy is measured in terms of hits/strip/event
where the integration time of an ``event'' is about 8 $\mu$s in the SVX.
The agreement 
here is good, but is complicated by many confounding factors.  The occupancy
measured in the SVX is a combination of beam-related occupancy and
electronic noise effects.  
These electronic noise effects are measured
with no beam in the machine, but are very sensitive to many other
conditions.  
The electronic noise measured at zero beam current (0.009 hits/strip/event)
has been added in quadrature with the beam-gas simulation hits to
obtain the model prediction shown in the Figure.
We also compare the occupancy prediction and measurement as functions of
the detector tracking layer in Figure~\ref{fig:HEPovsl} for 300 mA
total current.  
The electronic noise at zero beam current is determined for each layer 
and combined with the beam-gas simulation hits as described above.
The agreement for the SVX (layers 1-3) is quite good --
within a factor of two -- while for the gas tracking devices (the remaining
layers) is within measurement errors.

	Note that in all these comparisons between the measured
beam induced backgrounds
and the simulation prediction, 
the contribution of synchrotron radiation is
neglected.  The effects of synchrotron radiation 
are not expected to penetrate beyond the second tracking layer and thus
would only be observable in the radiation level and the first two layers
of the SVX.  In any case the simulation predicts that the contribution
to SVX detector occupancy
from synchrotron radiation during HEP running is 
negligible.

\subsection{Synchrotron Radiation}
Under normal HEP running the SR contribution to detector backgrounds
cannot be detected.  As mentioned above, simulation predicts the radiation
dose from SR to be 8\% that of lost particles, and the contribution to
detector occupancy only 1\%.  Therefore, to observe SR or to compare with
simulations it is necessary to adjust the orbit through the
IR so that the beam is intentionally 
placed far off-axis in the IR quadrupoles in order to 
generate large SR fluxes.  

As a machine studies experiment large SR fluxes were generated in two ways.
In the first (case A), the crossing angle was increased
from 2.0 mrad to 2.7 mrad (yielding a displacement of 19mm in Q1), 
and in the second (case B), a large amplitude bump
was added to a reduced pretzel to yield $x_{IP}^{\prime} = -1.9$ mrad and
$x_{IP} = $ 2.7 mm (yielding a displacement of 24 mm  in Q1).
For each measurement a single 10 mA bunch was used.
The tunes were adjusted as needed to maintain good lifetime ($\tau > 
1000$ min).  In addition, the CsI radiation detectors were monitored to
ensure that the particle backgrounds were maintained at a negligible level.
  The crossing angle and bump amplitude (needed for input to the
SR background simulation) were obtained by fitting the orbits
to separator kicks and magnetic corrector strengths respectively.

\begin{figure}[tbh]
\centering
\epsfig{file=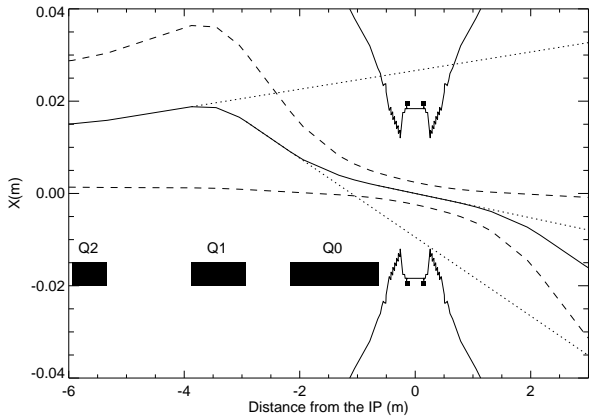, width=82.5mm}
\caption{Trajectory for the synchrotron radiation machine studies experiment
(case A).  The interaction point is located at (0,0) on this plot.  The
beampipe radiation monitors are shown as black squares located at
($\pm 0.13, \pm 0.02$).  The ``outside'' of CESR is $+x$ and the ``inside''
is $-x$.}
\label{fig:srtraj}
\end{figure}
The trajectory for case A is shown in Figure \ref{fig:srtraj}.
The incoming beam is bent in the strong
horizontally focusing quadrupole Q1. The radiation fan from the central
trajectory is shown by the upper and lower dotted lines.  The beam is
bent in Q0 (vertically focussing) and its radiation fan is shown by the 
lower and middle dotted lines.  Thus, the mask tips on the outside
of CESR are illuminated by Q1 radiation (from the core of the beam) 
and the downstream inside 
backscattering mask surface is illuminated by both Q1 and Q0 radiation
(from beamsize).  

\begin{figure}[tbh]
\centering
\epsfig{file=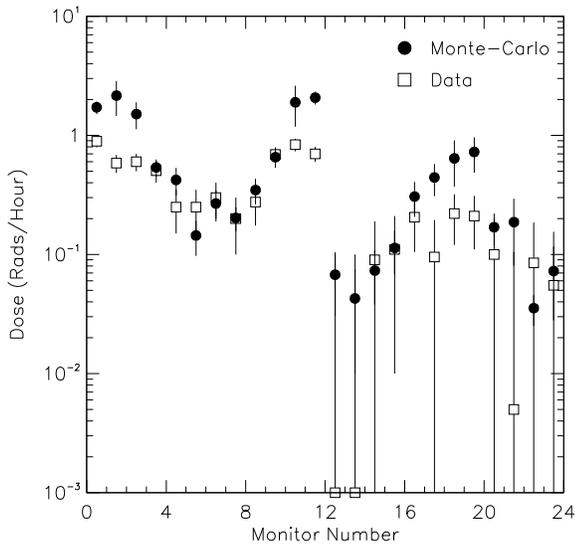, width=82.5mm}
\caption{Beampipe radiation monitor response for the synchrotron radiation
experiment labelled ``case A.'' The monitors 0-11 are ``downstream'' in the
experiment ($z>0$ in Figure \ref{fig:srtraj}), and 12-23 are
``upstream'' ($z<0$).  The numbering is as follows: 0 and 12 are located
in the plane of the machine on the {\em outside} of CESR, 
3 and 15 are located on
the top, 6 and 18 are located on the {\em inside} of CESR and 9 and 21 are
located on the bottom.}
\label{fig:srcase_a}
\end{figure}
The measured radiation distribution is compared to the Monte-Carlo simulation
results in Figure \ref{fig:srcase_a}.  
The highest radiation is observed in
monitors 0-11 (downstream; $z>0$ in Figure \ref{fig:srtraj}) and arises from
SR which backscatters from the downstream mask surface.  The monitors
on the downstream end (0-11) have a larger solid angle from the mask surface than
those on the upstream end (12-23) and those lying in the plane outside
of CESR have the largest solid angle and therefore the largest radiation dose
from $x$-ray fluorescence.  The radiation on the upstream monitors is due
primarily to tipscattering.  The tips on the outside of CESR are illuminated
by Q1 radiation.  The upstream monitors therefore have the largest solid angle
from the tip, and those on the inside have a larger solid angle than those on
the outside.  

\begin{figure}[tbh]
\centering
\epsfig{file=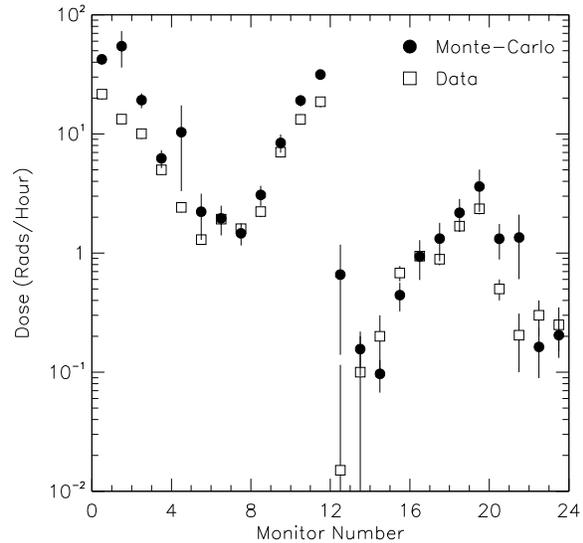, width=82.5mm}
\caption{Beampipe radiation monitor response for the synchrotron
radiation experiment labelled ``case B.''}
\label{fig:srcase_b}
\end{figure}

The comparison for case B is shown in Figure \ref{fig:srcase_b}.
 The trajectory is displaced
even further in Q1, so event greater SR fluxes are generated.  The radiation 
distribution is similar to that of the previous case.  The agreement between 
the SR simulation and the data is quite good - typically within a factor of
two.  The experimental cases each test two different aspects of the SR
simulation since the backscattering and tipscattering sources
separately give rise to radiation in the downstream and upstream monitors
respectively.  Therefore, agreement for monitors 0-11 can be viewed as a 
test of 
the backscattering portion of the simulation while agreement for 
monitors 12-23
tests the tipscattering portion of the simulation.

\section{Conclusion}

The CESR/CLEO II.V interaction region has been described.
The radiation dose measured by the IR radiation monitors during
172 weeks of operation is 22.2 krads, 12 krads of which was
accumulated during HEP data-taking.  Detailed comparisons of the
background simulation with measured detector background and
radiation dose have been performed.  The agreement between
measured and predicted radiation doses in the IR from
beam-gas scattering is quite good, within $\sim$25\% for the
various experiments described.  The comparison between beam-gas simulation
and measured detector occupancies is in many ways more
challenging.  The agreement here is within a factor of two over a range
in beam currents and better at higher currents.  Synchrotron
radiation is not readily observed during HEP running.  The results
of an experiment to generate a large SR dose were described and
a comparison to simulation shows agreement within a factor of two in
general, and much better in some cases.  We are confident in the
ability of these simulations to predict detector backgrounds
at CESR and have based the design of the CESR Phase III IR shielding
on them.

\end{document}